\documentclass[letterpaper]{article}

\usepackage{amsfonts,psfrag} 
\usepackage{amsmath, graphicx, epsfig}

\newcommand{\btheta}{\boldsymbol{\theta}}

\newcommand{\nr}{n \rightarrow \infty}

\newcommand{\eqd}{\stackrel{D}{=}}
\newcommand{\ez}{\mathbb{E}}
\newcommand{\pz}{\mathbb{P}}

\numberwithin{equation}{section}

\begin{document}
\title{Optimal scaling of the independence sampler: Theory and Practice}
\author{Clement Lee (Lancaster University -- currently at Newcastle University) \\ and Peter Neal$^\ast$ (Lancaster University)}
\maketitle

{\it Correspondence address:} Department of Mathematics and Statistics, Fylde College, Lancaster University, Lancaster, LA1 4YF, UK

{\it Running title:} Optimal scaling of the independence sampler

\begin{abstract}
The independence sampler is one of the most commonly used MCMC algorithms usually as a component of a Metropolis-within-Gibbs algorithm. The common focus for the independence sampler is on the choice of proposal distribution to obtain an as high as possible acceptance rate. In this paper we have a somewhat different focus concentrating on the use of the independence sampler for updating augmented data in a Bayesian framework where a natural proposal distribution for the independence sampler exists. Thus we concentrate on the proportion of the augmented data to update to optimise the independence sampler. Generic guidelines for optimising the independence sampler are obtained for independent and identically distributed product densities mirroring findings for the random walk Metropolis algorithm. The generic guidelines are shown to be informative beyond the narrow confines of idealised product densities in two epidemic examples.
\end{abstract}


\section{Introduction} \label{S:Intro}

The independence sampler is the incorporation of rejection sampling within an MCMC framework. The rejection sampler obtains samples from a random variable, $X$, with probability density function $f(\cdot)$ by first proposing a candidate value $y$ from a random variable, $Y$, with probability density function $q(\cdot)$, and secondly accepting $y$ as a sample from $X$ with probability $f(y)/\{K q(y)\}$, where $K = \sup_x f(x)/q(x)$. Otherwise $y$ is rejected, see \cite{Ripley}, page 60. The success of the rejection sampler depends upon making a good choice of $q(\cdot)$ such that $K (\geq 1)$ is small and that $q(\cdot)$ is straightforward to sample from. The MCMC independence sampler is the modification of the above where a Markov chain $X_0, X_1, \ldots$ is constructed with at iteration $t$, a candidate $y$ proposed from $Y$ and if accepted $X_t$ is set equal to $y$. Otherwise $X_t = X_{t-1}$. The rejection sampler, and consequently, the independence sampler can usually be implemented in a straightforward and efficient manner for low dimensional (target) distributions but as the dimension of $X$ increases it becomes increasingly more challenging to obtain a good choice of $q(\cdot)$. Therefore the independence sampler is rarely used as an MCMC algorithm in its own right but instead independence sampler moves are often incorporated within Metropolis-within-Gibbs to effectively update low dimensional subsets of $X$, see \cite{DR13}, page 15.

The main focus for independence samplers has been to choose the proposal density $q(\cdot)$ so as to have an acceptance probability as close to 1 as possible. Whilst this makes intuitive sense, the aim of the current paper is to challenge the idea of aiming for an acceptance probability as close to 1 as possible within the context of using independence samplers for updating augmented data in MCMC algorithms. Specifically, we are interested in the Bayesian statistical problem of obtaining samples from the posterior distribution of the parameters $\btheta$ of a model given data $\mathbf{x}$, $\pi (\btheta | \mathbf{x})$ in the case where the likelihood, $\pi (\mathbf{x} | \btheta)$ is intractable. We assume that given augmented data $\mathbf{y}$, $\pi (\mathbf{y}, \mathbf{x} | \btheta)$ is tractable and an MCMC algorithm can be constructed to obtain samples from the joint posterior of $\btheta$ and $\mathbf{y}$, $\pi (\btheta, \mathbf{y} | \mathbf{x})$. Then it is natural to construct an MCMC algorithm which alternates between updating the parameters and the augmented data as follows:
\begin{enumerate}
\item Update $\btheta$ given $\mathbf{x}$ and $\mathbf{y}$. {\it
i.e.}~Use $\pi (\btheta | \mathbf{x}, \mathbf{y})$.
\item Update $\mathbf{y}$ given $\mathbf{x}$ and $\btheta$. {\it
i.e.}~Use $\pi (\mathbf{y} | \mathbf{x}, \btheta)$.
\end{enumerate}

Our focus is the use of independence samplers to update $\mathbf{y}$ given $\mathbf{x}$ and $\btheta$. For updating augmented data a natural independence sampler often presents itself. For example, in an epidemic modelling context  where $\mathbf{x}$ denotes the removal times of infected individuals, $\btheta$ denotes the infection and infectious period parameters and $\mathbf{y}$ denotes the infection times of individuals, a natural candidate for the infection time of individual $i$ who is removed at time $x_i$ is $y_i = x_i - D$, where $D$ denotes the infectious period distribution, see \cite{NR05}, \cite{XN14} and Section \ref{ss:homo}. For non-centered parameterisations, \cite{PRS}, we can often denote $\mathbf{Y}$ as a deterministic function $h (\btheta, \mathbf{U})$ with $\pi (\mathbf{x}| \mathbf{y}, \btheta)$ easy to compute, where $\mathbf{U}$ is a vector of independent and identically distributed uniform random variables, see \cite{NH15} and Section \ref{ss:BDM}. Then to update $U_i$ we can propose a new value from $U(0,1)$. The dimension of the augmented data, $\mathbf{y}$, can be orders of magnitude higher than $\btheta$ and $\mathbf{x}$, so updating one component of $\mathbf{y}$ at a time can be prohibitive. Therefore we seek generic guidelines for updating multiple components of $\mathbf{y}$ at a time and optimising the performance of the resulting independent sampler. Specifically, this work formalises findings in \cite{XN14} and \cite{NH15} in using the independence sampler for data augmentation giving simple guidelines for producing close to optimal independence samplers. The guidelines obtained are similar to those given in \cite{RGG97} for the random walk Metropolis algorithm and comparisons with the random walk Metropolis algorithm are made.

The paper is structured as follows. In Section \ref{S:Theo}, we study the properties of the independence sampler for  independent and identically distributed product densities $\pi (\mathbf{x}) = \prod_{i=1}^n f (x_i)$. This idealised scenario mimics the set up in \cite{RGG97} where optimal scaling of the random walk Metropolis algorithm was first explored and as in \cite{RGG97} allows us to get a handle on understanding the key factors in optimising the independence sampler. In particular, we show that the optimal number of components, $k$, of $\mathbf{x}$ to update, is the $k$ which maximises the mean number of components per move. In the case where this optimal $k$ is large this corresponds to a mean acceptance rate of approximately $23.4 \%$. Thus there is a somewhat surprising link with the optimal scaling of the random walk Metropolis algorithm, \cite{RGG97} with which we make comparison and highlight the benefits of the independence sampler. In Section \ref{S:Ex}, we explore the optimal performance of the independence sampler for increasingly complex problems. In Section \ref{S:Ex:Intro}, we study product Gaussian target densities with Gaussian and $t$-distribution proposals demonstrating the optimal scaling results obtained in Section \ref{S:Theo}. In Sections \ref{ss:homo} and \ref{ss:BDM} we apply  the independence sampler to two epidemic models, the classic homogeneously mixing SIR epidemic model,  \cite{Bailey} and \cite{OR99} and a birth-death-mutation (BDM) model for an emerging, evolving disease, \cite{Tanaka} and \cite{FearnheadPrangle}. In Section \ref{ss:homo}, we show that for the homogeneously mixing SIR epidemic model updating a proportion of the infection times so as to obtain a mean acceptance rate of approximately $23.4 \%$ is optimal. This demonstrates that as observed with the random walk Metropolis algorithm the findings of Section  \ref{S:Theo} are informative in designing independence samplers beyond the limited confines of product densities. For the BDM model in Section \ref{ss:BDM} the findings are somewhat different with a lower optimal mean acceptance rate corresponding to large scale data augmentation.  Finally, in Section \ref{S:Conc}, we make some concluding remarks highlighting the possible benefits of the independence sampler over random walk Metropolis for large scale data augmentation and the differences seen between the two epidemic models in Sections \ref{ss:homo} and \ref{ss:BDM}.





\section{Theoretical properties of the independent sampler} \label{S:Theo}

In this Section we consider the theoretical properties of the independence sampler for the special case where $\pi_n (\mathbf{x}^n) = \prod_{i=1}^n f(x_i)$, a product of independent and identically distributed univariate densities, $f(x)$. The main focus is on the asymptotic behaviour as the number of components, $\nr$ mirroring analysis performed in \cite{RGG97} for the random walk Metropolis algorithm. The aim is to characterise the optimal performance of the independence sampler in terms of the number of components to update and to draw interesting comparisons of similarities and differences with the random walk Metropolis algorithm.

For the independence sampler we propose to select uniformly at random $k$ components $\{I_1, I_2, \ldots, I_k \}$ from $\{1,2,\ldots, n\}$ to update. For $j \in \{I_1, I_2, \ldots, I_k \}$, $y_j$ is drawn from $Y$ with probability density function $q(y)$, whilst for $l \not\in \{I_1, I_2, \ldots, I_k \}$, $y_l = x_l$. Therefore the acceptance probability for the proposed move from $\mathbf{x}^n$ to $\mathbf{y}^n$ is
\begin{eqnarray} \label{eq:acc1}
\min \left\{ 1, \frac{\pi_n (\mathbf{y}^n)}{\pi_n (\mathbf{x}^n)} \times \frac{q (\mathbf{y}^n \rightarrow \mathbf{x}^n)}{q (\mathbf{x}^n \rightarrow \mathbf{y}^n)}  \right\} = \min \left\{ 1, \prod_{j=1}^k \frac{f (y_{I_j})/q(y_{I_j})}{f (x_{I_j})/q(x_{I_j})}  \right\}.
\end{eqnarray}
For $n=1,2,\ldots$ and $t=0,1,\ldots$, let $\mathbf{X}_t^n = (X_{t,1}^n, X_{t,2}^n, \ldots, X_{t,n}^n)$ denote the position of the Markov chain after $t$ iterations. As in \cite{RGG97}, we assume that the Markov chain is initiated with $\mathbf{X}_0^n$ drawn from $\pi_n (\cdot)$ and thus for all $t \geq 0$, $\mathbf{X}_t^n \sim \pi_n (\cdot)$. The independent and identically distributed nature of the stationary and proposal distributions means that as in \cite{RGG97} it suffices to focus on the behaviour and performance of the independence sampler on the first component only. Specifically, for $t \geq 0$, letting $\mathbf{Z}^n_t = \mathbf{X}^n_{[nt]}$ we show that for fixed $k$, as $\nr$, the movement in the first component of $\mathbf{Z}^n_t$ converges to a Markov jump process with jumps governed by $f (\cdot)$ and $q(\cdot)$.

Let $\omega (x) = f(x)/q(x)$, then for the independence sampler to be well-behaved we require that $\sup_x \omega (x) < \infty$, see \cite{Tiernay} and we make this assumption throughout. For a move to occur in the first component we must propose to move the first component and $k-1$ other components from $\{2,3, \ldots, n\}$. Let $\{J_1, J_2, \ldots, J_{k-1} \}$ be a random sample from $\{2,3,\ldots, n\}$ with $W_{k-1} (\mathbf{x}^{n-}) = \prod_{i=1}^{k-1} \omega (Y_{J_i})/\omega (x_{J_i})$, where $\mathbf{x}^{n-} = (x_2, x_3, \ldots x_n)$. Define $\mathbf{Y}^{n-}$, $\mathbf{X}^{n-}$ and $\mathbf{y}^{n-}$ in the obvious fashion. Then we define
\begin{eqnarray} \label{eq:acc2}
H (y, \mathbf{x}^n) &=& H(y, x_1, \mathbf{x}^{n-}) \nonumber \\ &=& \ez_{\mathbf{Y}^{n-}, \mathbf{J}_{k-1}} \left[ 1 \wedge \frac{\omega (y)}{\omega (x_1)} W_{k-1} (\mathbf{x}^{n-}) \right] \nonumber \\
&=& \ez_{\mathbf{Y}^{n-}, \mathbf{J}_{k-1}} \left[ 1 \wedge \frac{\omega (y)}{\omega (x_1)} \prod_{i=1}^{k-1} \frac{\omega (Y_{J_i})}{\omega (x_{J_i})} \right],
\end{eqnarray}
where $\mathbf{J}_{k-1} = (J_1, J_2, \ldots, J_{k-1})$.
A useful observation is that the proposed values $(Y_1, Y_{J_1}, \ldots, Y_{J_{k-1}})$ are independent of $\mathbf{x}^n$. Let $H^\ast (y, x_1) = \ez_{\mathbf{X}^{n-}} [H (y, x_1,\mathbf{X}^{n-})]$ and let
\begin{eqnarray} \label{eq:acc3}
\mathcal{A}_n = \left\{ \mathbf{x}^n; \int | H (y, \mathbf{x}^n) - H^\ast (y,x_1)| q(y) \,dy \leq n^{-1/8} \right\}
\end{eqnarray}
We have the following Lemma which mirrors \cite{RGG97}, Lemma 2.1, which states that with sufficiently high probability we can focus upon $\mathbf{X}_{[nt]}^n$ ($\mathbf{Z}_t^n$) contained in $\mathcal{A}_n$. The proof of Lemma \ref{Lemma1} is given in appendix \ref{app:A}.
\newtheorem{theorem}{Lemma}[section]
\begin{theorem} \label{Lemma1} For $t >0$,
\begin{eqnarray} \label{eq:acc4}
\pz ( \mathbf{Z}^n_s \in \mathcal{A}_n, 0 \leq s \leq t) \rightarrow 1 \hspace{0.5cm} \mbox{as } \nr.
\end{eqnarray}
\end{theorem}

We are now in position to state and prove the main result of this Section, Theorem \ref{Thm}.
\newtheorem{thm_main}[theorem]{Theorem}
\begin{thm_main} \label{Thm} For $k \in \mathbb{N}$, let $\mathbf{X}_0^n \sim \pi_n$, then
\begin{eqnarray} \label{eq:gen:T1} Z^n_{\cdot,1} \Rightarrow Z_\cdot \hspace{0.5cm} \mbox{as } \nr, \end{eqnarray} where
$Z_\cdot$ is a Markov jump process with infinitesimal generator
\begin{eqnarray} \label{eq:gen:T2}
G h (x) &=& k \int \{ h (y) - h (x) \} H^\ast (y,x) q(y) \, dy, \end{eqnarray} for any $C_c^\infty$ function $h$.
\end{thm_main}
{\bf Proof.}
We begin by defining the (discrete time) generator of $\mathbf{X}^n$,
\begin{eqnarray} \label{eq:gen1}
G_n h (\mathbf{x}^n) = n \ez \left[ \{ h (\mathbf{Y}^n) - h (\mathbf{x}^n) \}  \left\{ 1 \wedge \frac{\pi_n (\mathbf{Y}^n)}{\pi_n (\mathbf{x}^n)} \right\} \right], \end{eqnarray}
where $h$ is any $C_c^\infty$ function of the first component.
Note that if there is no proposed update in the first component then $Y_1^n = x_1$. Therefore letting $\chi^n =1$ if there is a proposed update of the first component and 0 otherwise, we have that
\begin{eqnarray} \label{eq:gen2}
G_n h (\mathbf{x}^n) &=& \sum_{i=0}^1 n \pz (\chi^n = i) \ez \left[ \left. \{ h (\mathbf{Y}^n) - h (\mathbf{x}^n) \}  \left\{ 1 \wedge \frac{\pi_n (\mathbf{Y}^n)}{\pi_n (\mathbf{x}^n)} \right\} \right| \chi^n =i \right] \nonumber \\
&=& n \times \frac{k}{n} \times \ez \left[\left. \{ h (\mathbf{Y}^n) - h (\mathbf{x}^n) \}  \left\{ 1 \wedge \frac{\pi_n (\mathbf{Y}^n)}{\pi_n (\mathbf{x}^n)} \right\} \right| \chi^n =1 \right] \nonumber \\
&=& k \ez_{Y_1} \left[ (h (Y_1) - h (x_1)) \ez_{\mathbf{Y}^{n-}, \mathbf{J}_{k-1}} \left[ 1 \wedge \frac{\omega (Y_1)}{\omega (x_1)} \prod_{j=1}^{k-1} \frac{\omega (Y_{J_j})}{\omega (x_{J_j})} \right] \right]. \nonumber \\ \end{eqnarray}

We compare $G_n h (\mathbf{x}^n)$ with the generator $G h(x)$ defined in \eqref{eq:gen:T2} for the limiting jump process. Now by \eqref{eq:acc3},
 for all $\mathbf{x}^n \in \mathcal{A}_n$ and $h \in C_c^\infty$,
\begin{eqnarray} \label{eq:gen4}
&& |G_n h (\mathbf{x}^n) - G h (x_1)| \nonumber \\ &=& \left|  \int \{ h (y) - h (x_1) \} q(y) \left( \ez \left[ 1 \wedge \frac{\omega (y)}{\omega (x_1)} \prod_{j=1}^{k-1} \frac{\omega (Y_{J_j})}{\omega (x_{J_j})} \right] -
H^\ast (y,x)  \right) \, dy  \right| \nonumber \\
&=& \left|  \int \{ h (y) - h (x_1) \} q(y) \left( H(y, x_1, \mathbf{x}^{n-}) -
H^\ast (y,x)  \right) \, dy  \right| \nonumber \\
& \leq & 2 \sup_z | h(z)| \int q(y) \left( H(y, \mathbf{x}^n) -
H^\ast (y,x)  \right) \, dy \nonumber \\
& \leq & 2 \sup_z | h(z)| n^{-\frac{1}{8}} \rightarrow 0 \hspace{0.5cm} \mbox{as } \nr. \end{eqnarray}
Hence,
\begin{eqnarray} \label{eq:gen5}
\sup_{\mathbf{x}^n \in \mathcal{A}_n} | G_n h (\mathbf{x}^n) - G h (x_1) | \rightarrow 0 \hspace{0.5cm} \mbox{as } \nr.
\end{eqnarray}

The Theorem follows along identical lines to \cite{RGG97}, Theorem 1.1. Since $C_c^\infty$ separates points (see, \cite{EK}, page 113), the Theorem follows from \eqref{eq:gen5} and Lemma \ref{Lemma1} by Corollary 8.7 (f) of Chapter 4 of \cite{EK}. \hfill $\square$

We proceed by discussing properties of the limiting jump process. Let
\begin{eqnarray}
\label{eq:prop:1} W_k^\ast \eqd \prod_{i=1}^k \frac{\omega (Y_i)}{\omega (X_i)},
\end{eqnarray}
where $Y_i \sim q (\cdot)$ and $X_i \sim f (\cdot)$. Then $\ez [ 1 \wedge W_k^\ast]$ denotes the mean acceptance probability, in stationarity, of a proposed move
and $k \ez [ 1 \wedge W_k^\ast]$ denotes the corresponding mean number of components updated. Moreover, $k \ez [ 1 \wedge W_k^\ast]$ denotes the mean number of jumps, per unit time, of the limiting jump process, and hence, we seek $k$ which maximises $k \ez [ 1 \wedge W_k^\ast]$.

The distribution of $W_k^\ast$ depends largely on the {\it closeness} of the target ($f(\cdot)$) and proposal ($q(\cdot)$) distributions with $W_k^\ast \equiv 1$ if for all $x$, $f(x)\equiv q(x)$. Let $g(x) = \log \omega (x) =  \log f(x) - \log q(x)$, then
\begin{eqnarray}
\label{eq:prop:2} \log W_k^\ast \eqd \sum_{i=1}^k \{ g(Y_i) - g (X_i) \},
\end{eqnarray}
where the $\{ g(Y_i) - g (X_i) \}$ are independent and identically distributed. Note that $\ez [ g(Y_1)] = -D (q \| f)$ and
$\ez [ g(X_1)] = D (f \| q)$, where for two probability density functions $u$ and $v$,
\begin{eqnarray}
\label{eq:prop:3a}
D (u \| v) = \int u(x) \log \{ u(x)/v(x) \} \, dx \end{eqnarray} is the Kullback-Leibler divergence. Hence, \begin{eqnarray}
\label{eq:prop:3b} \ez [  g(Y_1) - g (X_1)] = - \{ D (q \| f) + D (f \| q)\} = -I, \mbox{ say}, \end{eqnarray} which makes explicit the role played by the {\it closeness} of the two densities.
It should be noted that $I = \infty$ if there exists $x$ such that $q(x)>0$ and $f(x)=0$, in such cases efficient independence sampling may still exist, for example, $X \sim U(0,1)$ and $Y \sim U(0,1 + \epsilon)$ for small, positive $\epsilon$.

For finite $I$, it follows from \eqref{eq:prop:2} by the  Central limit Theorem that for large $k$, $\log W_k^\ast$ is approximately Gaussian with mean $k \ez [  g(Y_1) - g (X_1)]$ and variance $k var ( g(Y_1) - g (X_1)) = k J$, say. Now if $I$ is small, which will be the case where the Central limit theorem is relevant, then $q(x) \approx f(x)$. Moreover, if $f(x) = q(x) \{1 + \epsilon (x) \}$ where $\epsilon (x)$ is small, then it is straightforward to show that $I = \int q(x) \{\epsilon (x)^2 + O (\epsilon(x)^3\} \, dx$ and that  $J = 2 \int q(x) \{\epsilon (x)^2 + O (\epsilon(x)^3\} \, dx \approx 2 I$.
Thus for $k$ large, with $\log W_k^\ast \approx V_k^\ast \equiv N (- k I, kJ)$, we have by \cite{RGG97}, Proposition 2.4, that
\begin{eqnarray}
\label{eq:prop:4}
k \ez [1 \wedge \exp (\log W_k^\ast)] & \approx & k \ez [ 1 \wedge \exp (V_k^\ast)] \nonumber \\
&=& k \times \left\{\Phi \left( - \frac{k I}{\sqrt{k J}} \right) + \exp \left( - kI + \frac{kJ}{2} \right) \Phi \left( - \sqrt{kJ} + \frac{kI}{\sqrt{k J}} \right) \right\} \nonumber \\
&\approx& k \times 2 \Phi \left ( - \sqrt{\frac{k I}{2}} \right),
\end{eqnarray}
where the latter approximation follows from setting $J = 2I$. Replacing $k$ by $z^2$ and $I$ by $\tilde{I}=\sqrt{2} I$ in the right hand side of \eqref{eq:prop:4}, we obtain $j(z) = 2 z^2 \Phi (- z\sqrt{\tilde{I}}/2)$, which is the function maximized in \cite{RGG97}, Corollary 1.2 to maximise the optimal scaling of the random walk Metropolis algorithm. The only difference is the form of $I$ which here depends upon the Kullback-Leibler divergence between the target and proposal distribution, whereas in \cite{RGG97} $I \equiv \ez_f [ (f^\prime(X)/f(X))^2]$ and depends upon the {\it smoothness} of $f(\cdot)$. Most importantly, $z^2I =2.835$ maximises $j(z)$ and therefore $k$ should be chosen approximately equal to $2.835/I$. Thus if $I$ is small (there is close agreement between $f(\cdot)$ and $q(\cdot)$) $k$ will be large. Moreover, mirroring \cite{RGG97}, Corollary 1.2, such a $k$ corresponds to a mean acceptance probability of (approximately) $0.234$. Thus it is not necessary to compute $I$ but instead suffices to monitor the mean acceptance probability. This will be shown to be a useful guiding principle in the examples below. However, it should be noted that scenarios exist, see Section \ref{ss:homo} below, where the acceptance rate is above (below) 0.234 for all $k$, in such cases it is optimal to choose $k=n$ $(k=1)$.

Returning to optimising the independence sampler in the case $X \sim U(0,1)$ and $Y \sim U(0,1 + \epsilon)$, it is straightforward to show that the probability a proposed move is accepted is $(1+\epsilon)^{-k}$. Optimising the function $k (1+\epsilon)^{-k}$  gives $k = 1/\log (1 + \epsilon)$, and hence for small $\epsilon$, $k \approx 1/\epsilon$. Thus as $\epsilon \downarrow 0$, the optimal acceptance probability ($(1+\epsilon)^{-1/\log (1 + \epsilon)} \approx (1-\epsilon)^{1/\epsilon}$) converges to $\exp(-1)=0.368$. Therefore non-trivial asymptotic acceptance probabilities can exist in the case $I=\infty$ and typically these will be different from 0.234.


A key question is how does the independence sampler compare to the random walk Metropolis algorithm. Provided $\sup_x \omega (x) < \infty$, Theorem \ref{Thm} holds and we have that the mixing of the independence sampler algorithm is $O(n)$, the same order of mixing as for the random walk Metropolis algorithm for continuous (and sufficiently differentiable) densities. The mixing of the random walk Metropolis algorithm for discontinuous densities is $O(n^2)$, \cite{NRY12} whilst modifications such as Metropolis adjusted Langevin algorithms (MALA) and hybrid Monte Carlo (HMC) algorithms mix in $O(n^{\frac{1}{3}})$ and $O(n^{\frac{1}{4}})$ iterations, see \cite{RR98} and \cite{Beskos}, respectively, for sufficiently well behaved (continuous) target densities. Thus the independence sampler is competitive with the random walk Metropolis algorithm and Theorem \ref{Thm} holds under very weak conditions compared with those imposed for corresponding random walk Metropolis algorithms. The similarity of the right hand side of \eqref{eq:prop:4} to $j(z)$ might suggest that computing $I$ for the two algorithms would assist in comparing there performances with smaller $I$ the better. However, the different nature of the moves, global in the independence sampler and local in the random walk Metropolis, means that this is not the case.
In simulation studies with $X \sim N(0,1)$, $Y \sim N(0,\phi^2)$ and a range of $n \geq 50$, the independence sampler, with appropriately chosen $k$ was found to outperform the optimal random walk Metropolis algorithm ($\sigma = 2.4/\sqrt{n}$) for $1 \leq \phi \leq 2.4$.  Thus the independence sampler is competitive with, and often superior to, random walk Metropolis, for continuous target densities so long as a reasonable choice of $q(\cdot)$ is made, and is clearly preferable for discontinuous target densities which is often the case in real life Bayesian problems, see Section \ref{S:Ex}.

\section{Examples} \label{S:Ex}

\subsection{Introduction} \label{S:Ex:Intro}

In this Section we illustrate how large scale independence sampling can be exploited to construct effective MCMC algorithms.  We start with an independent and identically distributed Gaussian product density as the target distribution and consider both Gaussian and $t$-distribution proposals. Specifically, we take $\pi (\mathbf{x}) = \prod_{i=1}^n f (x_i)$, where $f (x)$ is a standard Gaussian density. The proposal distributions are symmetric about 0 with Gaussian proposals $q_N (y) = (\sqrt{2 \pi} \lambda)^{-1} \exp (-y^2/2 \lambda^2)$, where $\lambda \geq 1$ and $t$-distribution proposals $q_t (y) = \Gamma ((\nu +1)/2)/ (\sqrt{\nu \pi} \Gamma (\nu/2)) (1 + x^2/\nu)^{- \frac{\nu +1}{2}}$ $(\nu \in \mathbb{N})$. We conducted a simulation study using 5 Gaussian and 5 $t$-distribution proposals with $n=1000$ and $10^6$ iterations of the MCMC algorithm starting from the stationary distribution. For each proposal distribution we considered 50 choices of $k$, the exact choices of which depended on $I$ and were chosen to give acceptance rates on the full range 0 to 1.

\begin{figure} [h]
    \includegraphics[width = 6cm]{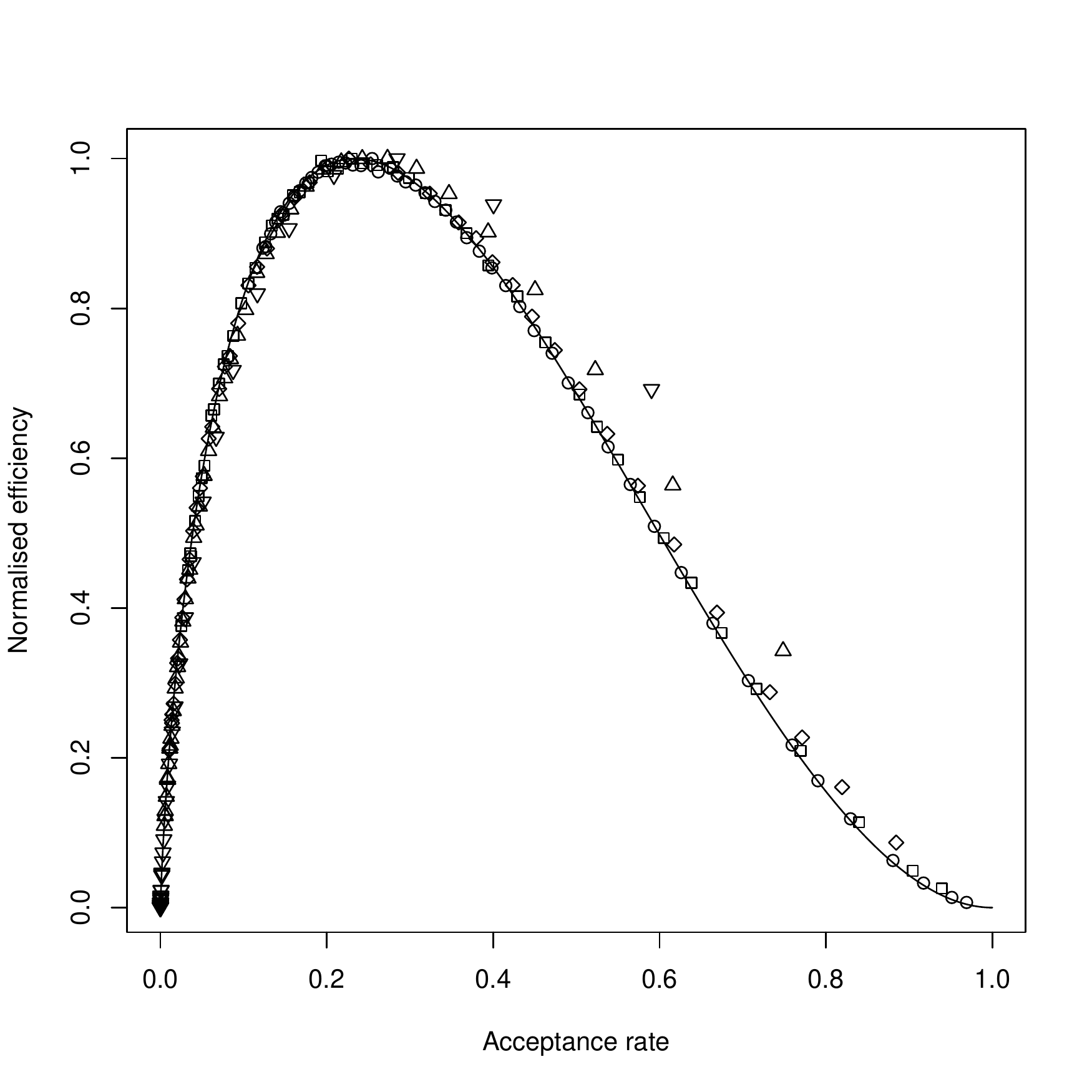}
		\includegraphics[width = 6cm]{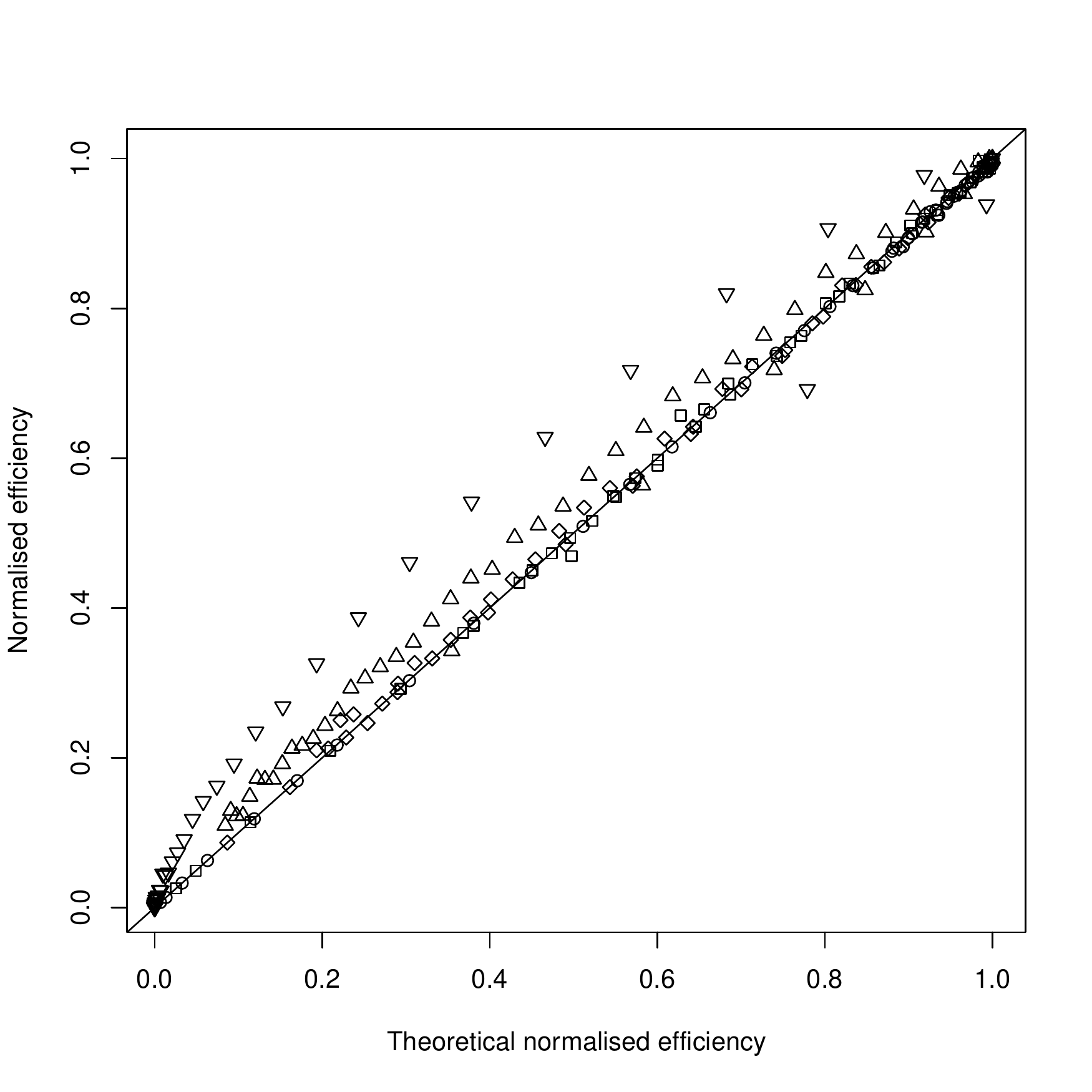}
  \caption{Gaussian proposal $\lambda =1.05 (\bigcirc), 1.1 (\square), 1.2 (\Diamond), 1.5 (\triangle), 2.0 (\bigtriangledown)$. (a) Solid line given by $\displaystyle j(z) = 2 z^2 \Phi (- z/2)/1.3257$ plotted against acceptance rate. (b) Solid line $x=y$.} \label{fig.gaus}
\end{figure}

For the Gaussian proposal it is straightforward to show that $I = 1/2 (\lambda - 1/\lambda)^2$. We considered $\lambda =1.05,1.1,1.2,1.5,2$ with corresponding $I =0.0048, 0.0182, 0.0672, 0.347, 1.125$. A key quantity for comparing the independence sampler for different choices of $\lambda$, and hence $I$, is the normalised efficiency. We define the normalised efficiency for $k$ as the mean number of components updated ($k \times$ acceptance rate) when proposing to update $k$ components divided through by the maximum mean number of components updated for $j=1,2, \ldots, n$. Correspondingly the normalised theoretical efficiency is given by $j (z) =2 z^2 \Phi (-z/2)/\sup_y \{2 y^2 \Phi (-y/2) \} = 2 z^2 \Phi (- z/2)/1.3257$ from applying the central limit theorem approximation obtained in Section \ref{S:Theo}. The plots in Figure \ref{fig.gaus} show that in all cases the optimal acceptance rate is close to 0.234 with very similar behaviour for the normalised efficiency varying with acceptance rate, even for $\lambda =2$ with $I=1.125$. Similar results are obtained in Section  in \cite{NR06}, Section 6 for the optimal performance of the random walk Metropolis algorithm. As $\lambda \downarrow 1$, $I \downarrow 0$ and the agreement between the observed normalised efficiency normalised theoretical efficiency becomes very close.

\begin{figure} [h]
    \includegraphics[width = 6cm]{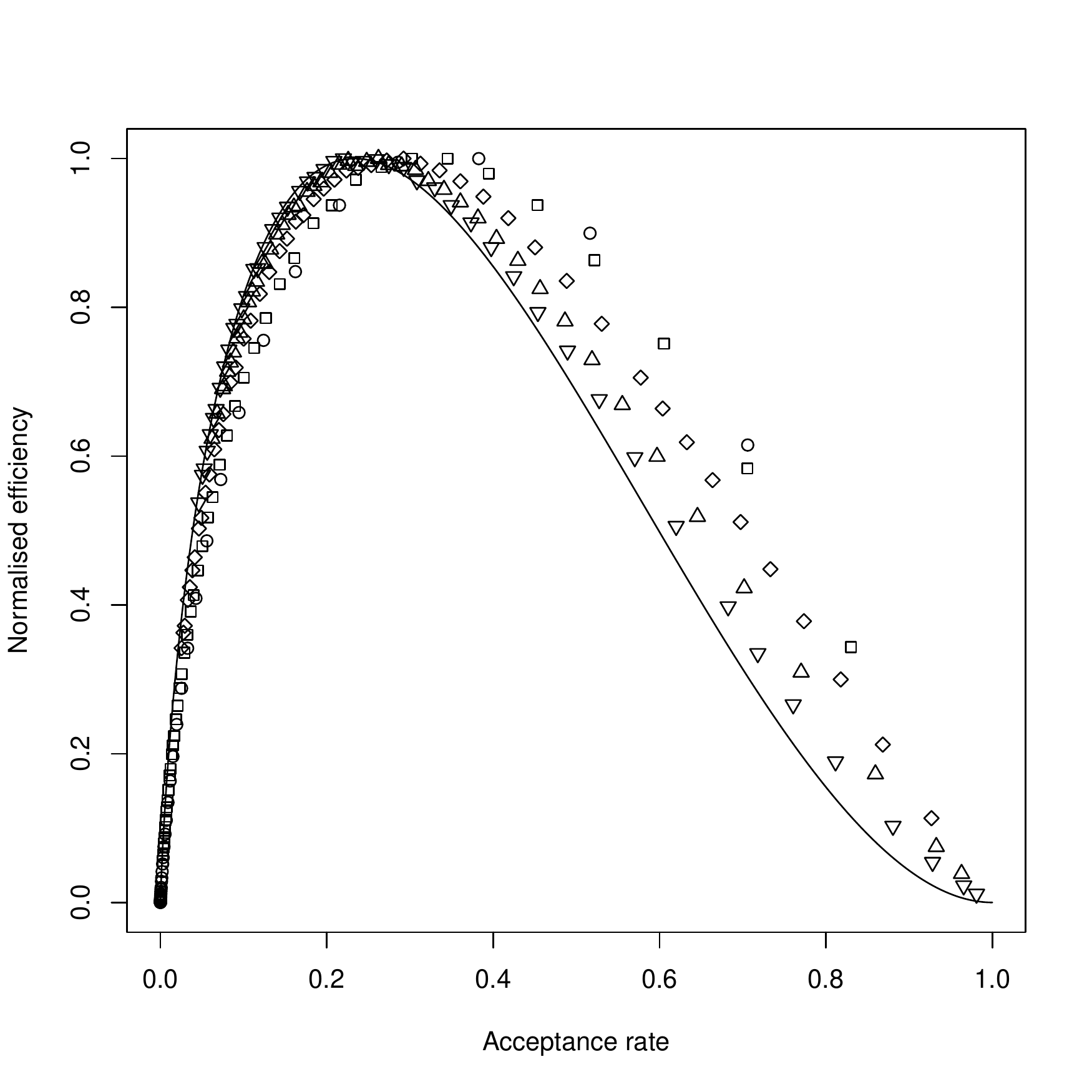}
		\includegraphics[width = 6cm]{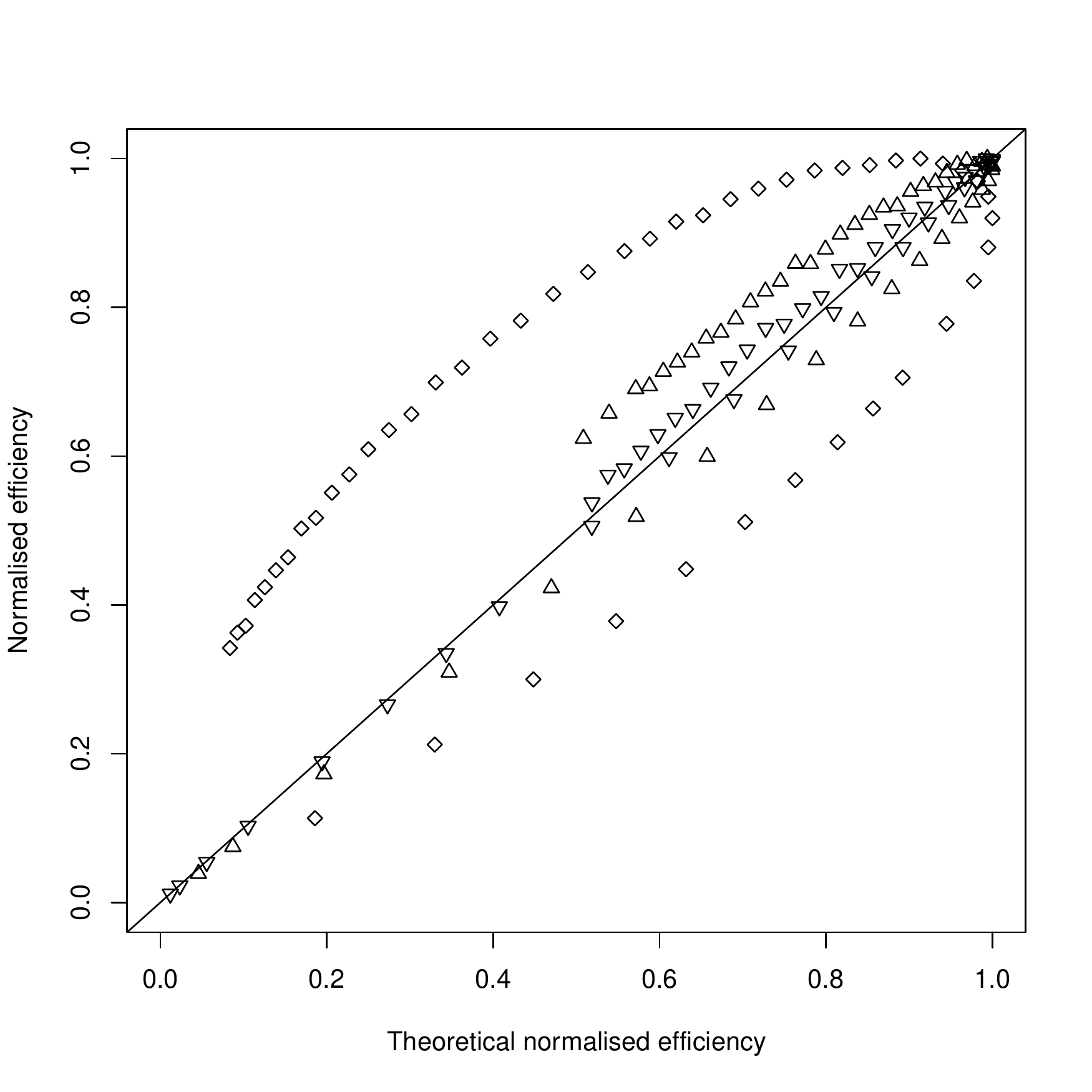}
  \caption{Gaussian proposal $t =1 (\bigcirc), 2 (\square), 5 (\Diamond), 10 (\triangle), 20 (\bigtriangledown)$. (a) Solid line given by $\displaystyle j(z) = 2 z^2 \Phi (- z/2)/1.3257$ plotted against acceptance rate. (b) Solid line $x=y$.} \label{fig.t}
\end{figure}

For the $t$-distribution, $I=\infty$ for $\nu=1,2$, otherwise
\[ I = \frac{1}{\nu-2} + \frac{\nu+1}{2} \left\{ \ez [ \log (1 + X^2/\nu)] - \ez [ \log (1 + Y_\nu^2/\nu)] \right\}, \] where $X \sim N(0,1)$ and $Y \sim t_\nu$. It is not possible to obtain a closed form analytical expression for $I$ but it is straightforward to estimate using Monte Carlo integration. We consider $\nu=1,2,5,10,20$ with corresponding $I = \infty, \infty, 0.1582, 0.0338,0.0083$. The plots in Figure \ref{fig.t} show that the optimal acceptance rate is higher than 0.234 for a $t$-distribution proposal with an optimal acceptance rate of 0.383 corresponding to $k=3$ for a $t_1$ proposal. Note that this is close to $\exp(-1)$, the optimal acceptance rate of the uniform distributions example given in Section \ref{S:Theo}. It is worth noting that choosing $k$ to obtain an acceptance rate of approximately 0.234 is in general a good approach as only a small loss in efficiency is observed. As $\nu$ increases the optimal acceptance rate converges towards 0.234 and the normalised efficiency tends towards the theoretical normalised efficiency given by the central limit theorem approximation. This is further demonstrated in Figure \ref{fig.t2} by plotting normalised efficiency against normalised theoretical efficiency. Note that $\nu=1$ and $\nu=2$ do not feature on this plot as $I =\infty$.

\subsection{Homogeneously mixing SIR epidemic} \label{ss:homo}

In this Section we show how the importance sampler can be applied to temporally observed, homogeneously mixing SIR epidemic models, \cite{Bailey,OR99}. We assume that there is a population of size $N$ with the disease introduced into the population via a single introductory case. (The extension to multiple introductory cases is trivial.) We assume that the disease follows an $SIR$ epidemic
model, where initially all individuals, except the introductory
case, are susceptible. On becoming infectious, an individual is
infectious for a given period of time, distributed according to a
Gamma random variable $Q \sim {\rm Gamma} (\alpha, \delta)$. (Alternative infectious period distributions can easily be considered.) Whilst infectious, an individual $i$, say, makes infectious contacts at the
points of a homogeneous Poisson point process with rate
$\beta$ with the individual contacted chosen uniformly at random from the entire population. Infectious contacts with susceptible individuals result in the immediate infection of the individual and the start of their infectious period. Infectious contacts with infectives have no effect on the recipient.

Suppose that $m$ individuals are infected during the course of the epidemic and we are analysing the completed epidemic data.
For each individual, $i$ say, infected during the course of the epidemic there will be an infection time, $I_i$ and a removal (recovery) time, $R_i$, which mark the start and end of the infectious period, respectively. We follow \cite{OR99}, \cite{NR05} and \cite{XN14} in assuming that the removal times, $\mathbf{R} = (R_1, \ldots, R_m)$ are observed, whilst the infection times $\mathbf{I} = (I_1, \ldots, I_m)$ are unobserved. Furthermore, we assume that the removal times are ordered such that $R_1 \leq R_2 \leq \ldots \leq R_m$. The key interest is in the posterior distribution of $\pi (\beta, \alpha, \delta | \mathbf{R})$ and to obtain samples from this distribution imputation of $\mathbf{I}$ is required.

\begin{figure} [h]
    \includegraphics[width = 6cm]{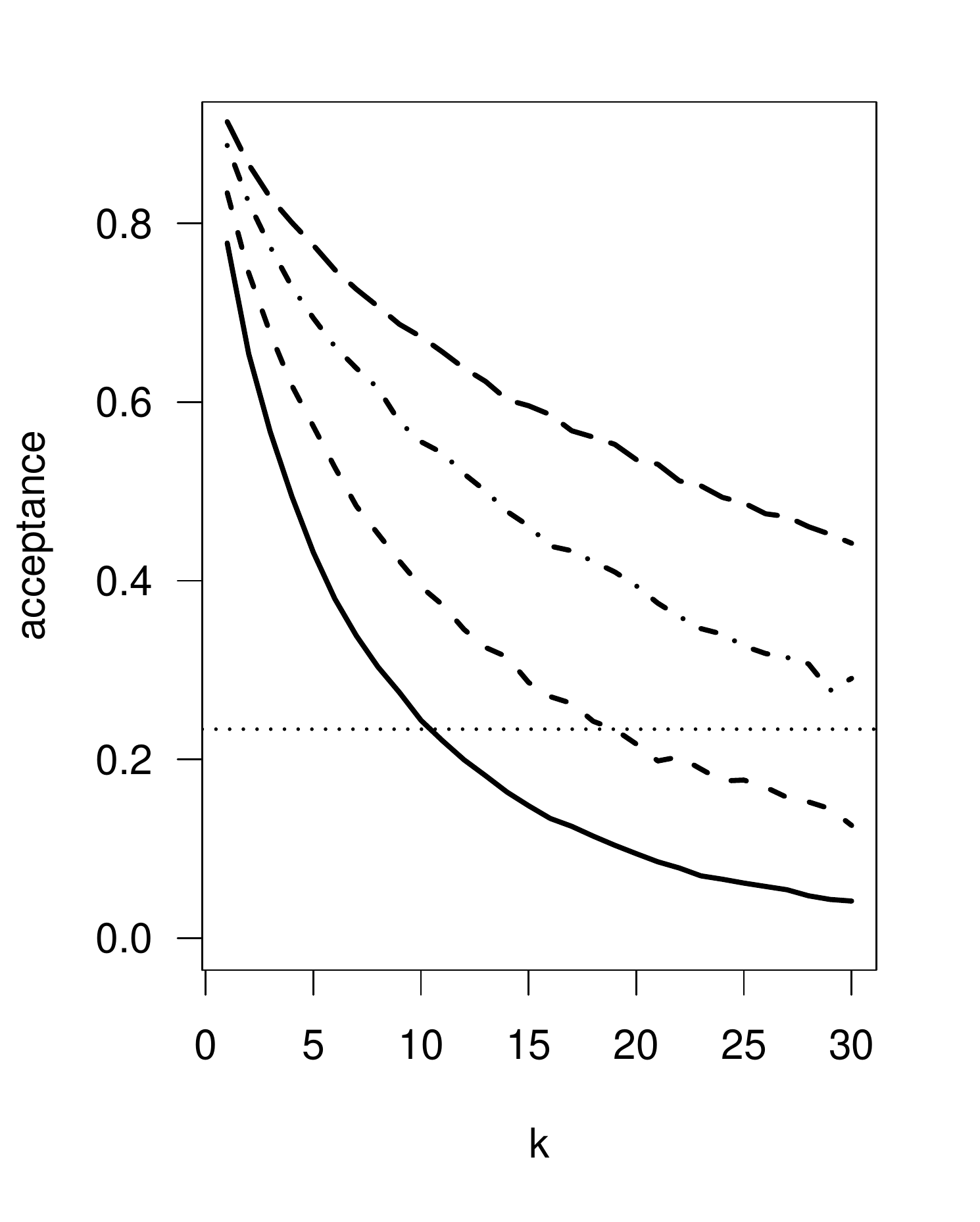}
		\includegraphics[width = 6cm]{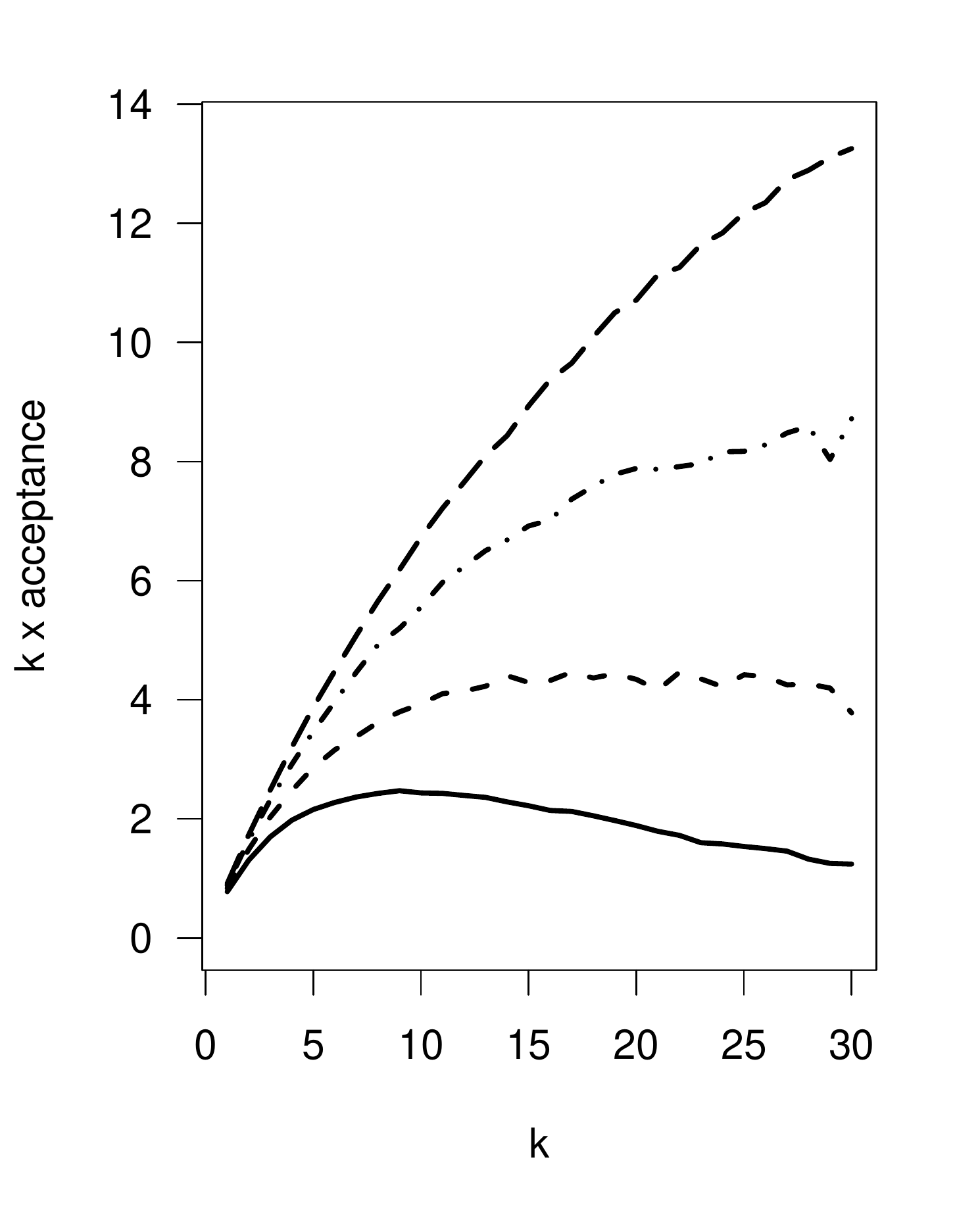}
  \caption{Acceptance rate (left) and mean number of components updated (right) against $k$ for $\alpha = 1$ (solid), $3$ (dashed), $
10$ (dot-dashed) and unknown (posterior mean 33.8) (long dashed).} \label{fig.hme_accept_kaccept_k_m030}
\end{figure}

We use the MCMC algorithm proposed in \cite{XN14}, Section 3 with the modification that the number of components to be updated is fixed to $k \in \{1,2 , \ldots, m\}$. As with \cite{XN14}, the MCMC algorithm is applied to the extensively studied Abakaliki smallpox outbreak, \cite[p.125]{Bailey}, \cite{OR99,OB01,McK14}, where $m=30$ and $N=120$. We considered various fixed values of $\alpha=1,3,10$ with optimal $k=9, 17$ and 30, respectively, based upon the maximised mean number of components updated over 100000 iterations, see Figure \ref{fig.hme_accept_kaccept_k_m030}. For $\alpha=1,3,10$, the corresponding values of $k$ which had acceptance rates closest to $23.4\%$ were $k=10,19$ and 29, respectively. Thus choosing $k$ so that the acceptance rate is close to $23.4\%$ is effective in obtaining a close to optimal algorithm. In \cite{XN14}, the situation where $\alpha$ is assumed to be unknown is also considered with the posterior mean of $\alpha$ being $33.8$. For unknown $\alpha$, the acceptance rate is above  $23.4\%$ for all $k$ and thus $k=m (=30)$ performs optimally.

\begin{figure} [h]
  \centering
  \vspace{-0.5cm}
  \includegraphics[width = 8cm]{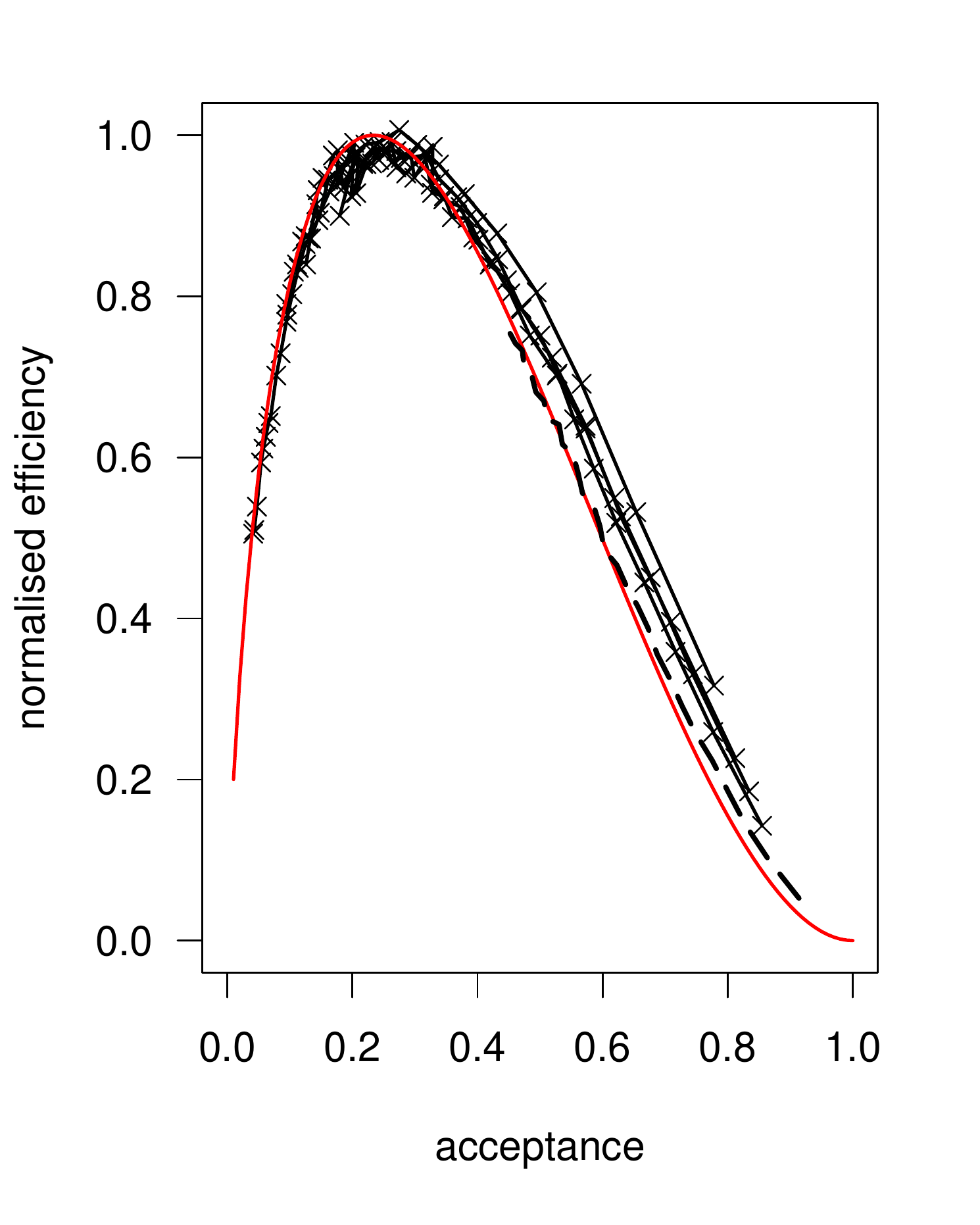}
  \caption{Normalised mean number of components updated against acceptance rate, overlaid by the theoretical normalised curve (red), given by $\displaystyle j(z) = 2 z^2 \Phi (- z/2)/1.3257$.} \label{fig.hme_kaccept_accept_m030}
\end{figure}

We can go further in illustrating the usefulness of the theoretical results derived in Section \ref{S:Theo} for choosing $k$. 
 In Figure \ref{fig.hme_kaccept_accept_m030}, we plot the normalised efficiency for $\alpha =1,2, \ldots, 9$, since for $\alpha >9$, the acceptance rate is always above $23.4\%$. Also on the plot (in red) is the normalised theoretical curve $j(z) = 2 z^2 \Phi (- z/2)/1.3257$ given by \eqref{eq:prop:4} against acceptance rate $2 \Phi (-z)$. In a similar fashion to  Section \ref{S:Ex:Intro} this illustrates that the asymptotic results which are valid as the number of components updated tend to $\infty$ are applicable for small $k$.

A simulation study was conducted to study the general applicability of the results obtained above for the Abakaliki data. Data sets were simulated with $N=200,400,600,800,1000,1200$, $m =0.25 N, 0.5 N, 0.75 N$ and $\alpha = 1,2,3,5,10,15,20$ with $\delta =0.1 \alpha$ chosen to give a mean infectious period of 10 and $\beta$ to give the mean size of a major epidemic outbreak to be 10. For each $\alpha$, the optimal $k$ increases with $N$ and vice versa. Throughout choosing $k$ with acceptance rate closest to $23.4\%$ produced close to optimal performance. Plots of the normalised efficiency against the acceptance rate showed increasing agreement with the asymptotic theoretical curve as $N$ increases.

\subsection{Birth-Death-Mutation model} \label{ss:BDM}

In this section we consider a birth-death-mutation (BDM)  model which is applicable to the early stages of a mutating disease. The model has previously been used by \cite{Tanaka,SFT,FearnheadPrangle,DMDJ,NH15} to analyse data from a tuberculosis outbreak in San Francisco in the early 1990s reported in \cite{Small}. We explore and seek to optimise the performance of the forward simulation MCMC algorithm introduced by \cite{NH15}. Note that all the other analyses reported above used ABC algorithms.

The  data consist of the genotypes of 473 bacteria samples sampled
from individuals infected with tuberculosis in San Francisco during
an observational period in 1991-92. The data are clustered by
genotype and summarised in Table \ref{tab:tub}. Let $N_t$ denote the
total number of tuberculosis cases at time $t$. The data are assumed
to be a random sample taken at time $T$, where $T = \min \{ t; N_t =
10000 \}$ evolving from $N_0 =1$.

\begin{table}
\caption{\label{tab:tub} Observed cluster size distribution of Tuberculosis bacteria genotype data, \cite{Small}.}
\begin{center}
\begin{tabular}{|l|rrrrrrrrrr|}
\hline Cluster size & 1 & 2 & 3&  4 & 5 & 8 & 10 & 15 & 23 & 30 \\
\hline Number of clusters & 282 & 20 & 13 & 4 & 2 & 1 & 1 & 1 & 1 & 1 \\
\hline
\end{tabular}
\end{center}
\end{table}

The BDM model is a Markov process defined as follows. Individuals are classified by (geno)type. Each individual born into the process has an exponentially distributed lifetime (infectious period) with mean $1/\delta$. Whilst alive individuals give birth (infects) and mutates at the points of independent homogeneous Poisson point processes with rates $\alpha$ and $\vartheta$, respectively. Each individual born inherits the (geno)type of their parent and all mutations result in the creation of a new, previously unseen (geno)type (infinite allele model, \cite{KC64}). We reparameterise the model by setting $\phi = \alpha + \delta + \vartheta$, $a = \alpha/\phi$ and $d = \delta/\phi$, where $\phi$ is the rate at which events occur for an individual, $a$ is the probability that the event is a birth (infection) and $d$ is the probability that the event is a death (recovery).  Since the stopping time $T$ at which the population is observed only depends upon the number of individuals alive in the population, there is no information in the data about $\phi$. Thus, without loss of generality, we assume $\phi =1$ making inference about $(a,d)$ given the genotype data $\mathbf{x}$. In order to construct a tractable likelihood it is necessary to generate the state of the population at time $T$, $N_T =10000$. This can be done using a non-centered parameterisation \cite{PRS} where the augmented data $\mathbf{y} = (\mathbf{u}, \mathbf{w}, \mathbf{v})$ consist of realisations of $U(0,1)$ with $(\mathbf{u}, \mathbf{w})$ combine with $(a,d)$ to generate the underlying state of the BDM model at time $T$ and $\mathbf{v}$ is used to estimate the probability of observing $\mathbf{x}$. Details of the construction are given in \cite{NH15}, Section 4.

The time consuming step of the MCMC algorithm for the BDM model is the simulation of the state of the process using $(\mathbf{u}, \mathbf{w})$ and $(a,d)$. In \cite{NH15}, $(a,d)$ are updated using random walk Metropolis keeping $(\mathbf{u}, \mathbf{w})$ fixed and $(\mathbf{u}, \mathbf{w})$ are updated using an independence sampler, draws from $U(0,1)$, keeping $(a,d)$ fixed. We thus focus on the independence sampler for updating $(\mathbf{u}, \mathbf{w})$. Note that $\mathbf{v}$ is updated by a separate independence sampler but this is very fast to implement (no need to simulate the BDM process), and so we don't comment on this step. The dimensions of $\mathbf{u}$ and $\mathbf{w}$ are the same but vary from iteration to iteration, typically being around 30000. To circumvent issues with this \cite{NH15} used random vectors of a fixed length $n=100000$ with only those elements needed to simulate the process used. In this paper we also used a fixed length vector updating $k$ out of $n$ components in $\mathbf{u}$ and $\mathbf{w}$ noting that in each simulation not all (updated) components will be used.

\begin{figure} [h]
    \includegraphics[width = 6cm]{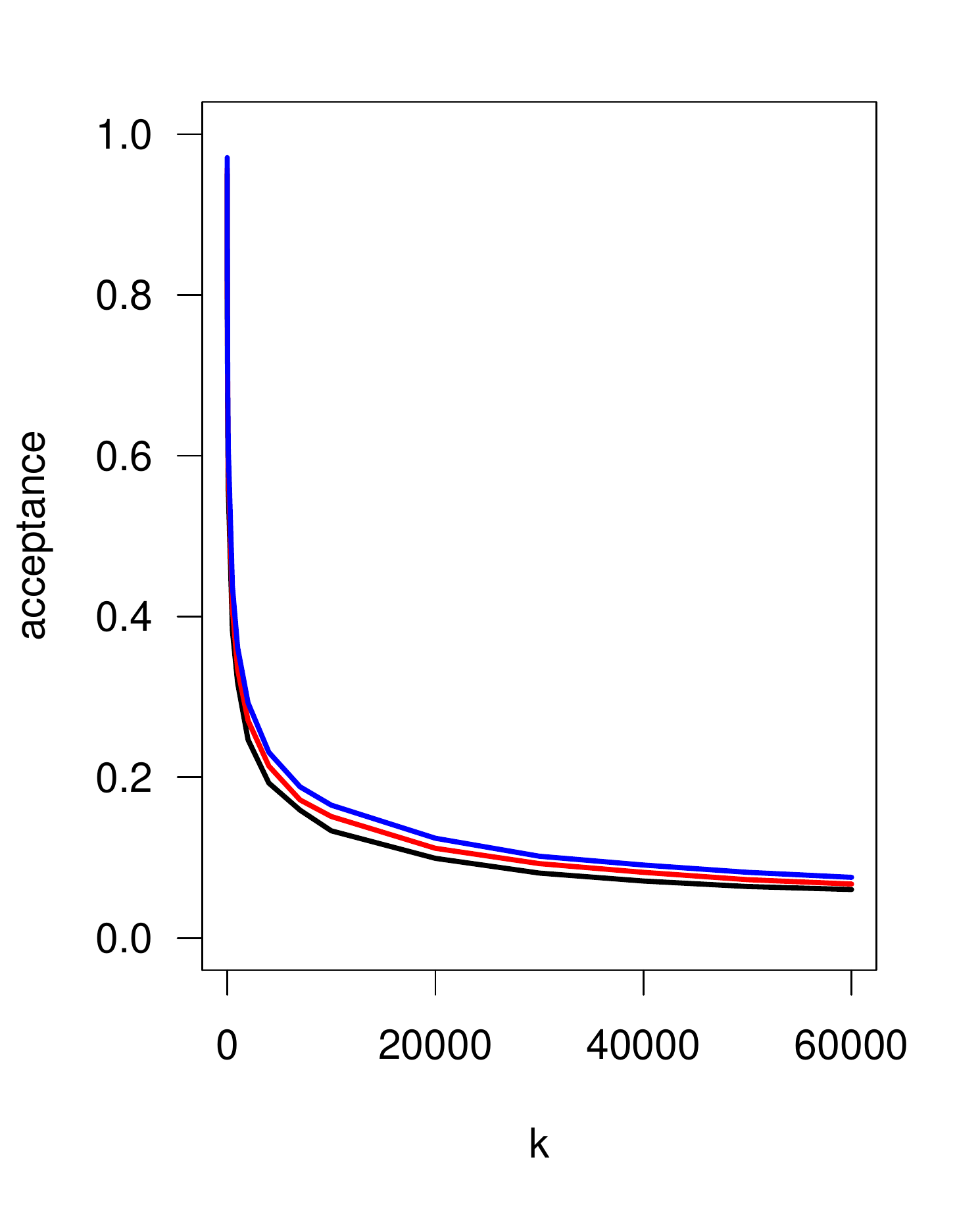}
		\includegraphics[width = 6cm]{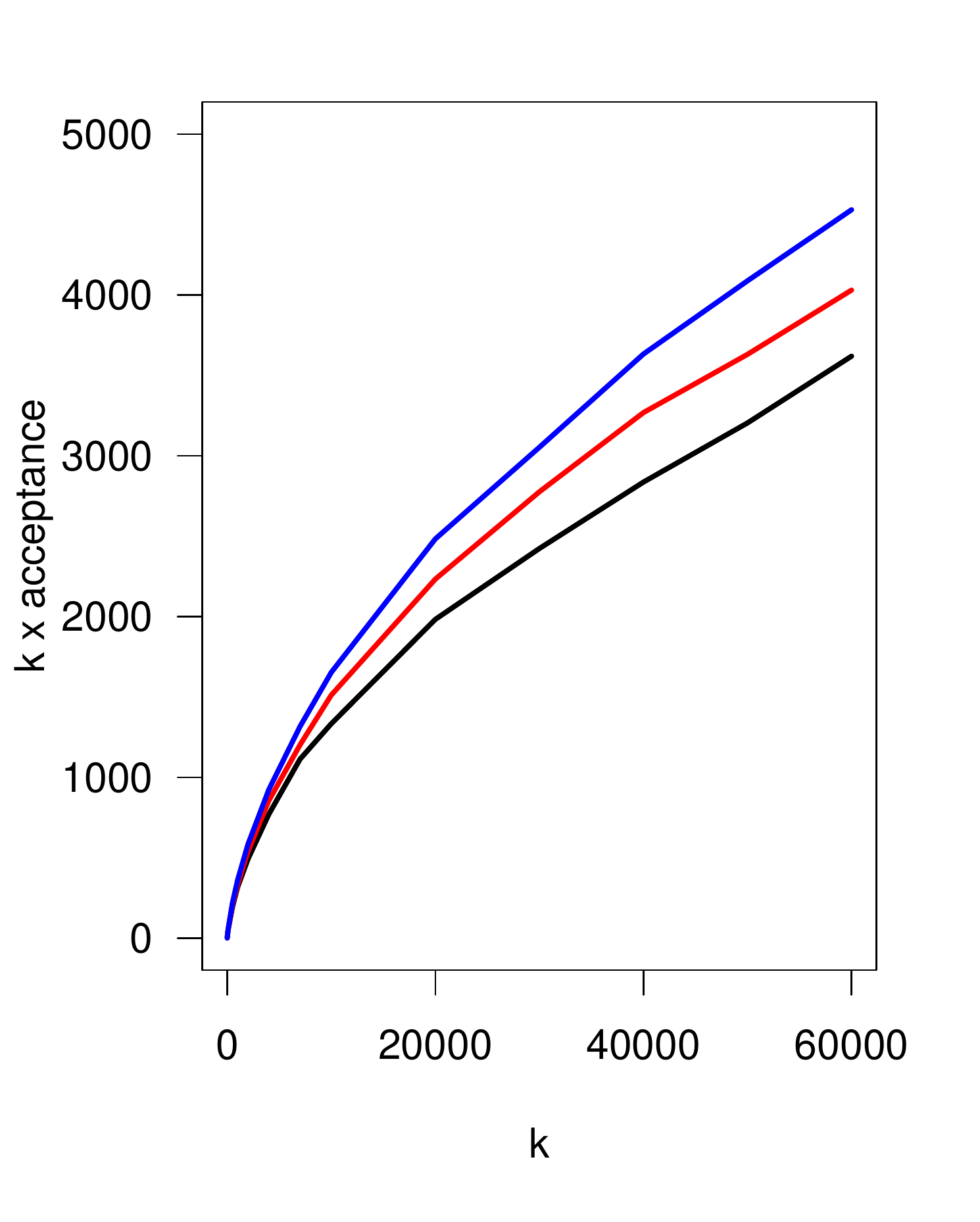}
  \caption{Acceptance rate (left) and mean number of components updated (right) against $k$ for $n = 60000$ (black), $80000$ (red) and $100000$ (blue).} \label{fig.bdm_accept_kaccept_k}
\end{figure}

In \cite{NH15}, $\mathbf{u}$ and $\mathbf{w}$ are broken down into blocks of 50 components with 1 component in each block proposed to be updated. This amounts to proposing to update $n/50=2000$ values in each iteration of which typically around 600 are used in the simulation.
In this paper we propose to update $k$ components each of $\mathbf{u}$ and $\mathbf{w}$, $(u_{I_1^u},u_{I_2^u},\ldots,u_{I_k^u})$ and $(w_{I_1^w},w_{I_2^w},\ldots,w_{I_k^w})$, where $\{I_1^u,I_2^u,\ldots,I_k^u\}$ ($\{I_1^w,I_2^w,\ldots,I_k^w\}$) is a uniformly random sample without replacement from $\{1,2,\ldots,n\}$, for the sake of consistency with the updating strategy throughout this paper. In addition to using different values for $k$, we also examine the performance of the algorithm using $n=60000,80000$ and the original $100000$, which are all found to be empirically sufficient. We ran the MCMC algorithm for $1.1\times 10^6$ iterations with the first $10^5$ iterations discarded as burn-in. The acceptance rate is plotted against $k$ for all three values of $n$ on the left of Figure \ref{fig.bdm_accept_kaccept_k}, which is analogous to Figure \ref{fig.hme_accept_kaccept_k_m030}, with the mean number of components updated on the right. The results shown in Figures \ref{fig.bdm_accept_kaccept_k} demonstrate an interesting departure from those found earlier in the paper with an optimal acceptance rate of $23.4\%$. The mean number of components updated increases with $k$ even as the acceptance rate drops below $5\%$. However, for both parameters $a$ and $d$, the effective sample size levels off at around 3000 for all $k \geq 2000$,  which suggests that seeking to optimise the mean number of components updated does not tell the full story in this case.

\section{Conclusions} \label{S:Conc}

In this paper we have demonstrated the potential benefits, both theoretical and practical, of the independence sampler over the random walk Metropolis algorithm. In particular, we have shown that simple choices of proposal distributions can be used to construct effective independence samplers and that similar considerations to the tuning of the random walk Metropolis algorithm are required. There are a number of points to consider in the wider application of the results derived in Section \ref{S:Theo} and applied in Section \ref{S:Ex}. Firstly, we have not considered the computational time required to update $k$ components. In the homogeneously mixing epidemic model (Section \ref{ss:homo}), and in particular, the BDM model (Section \ref{ss:BDM}) the time taken per iteration was essentially independent of $k$. However, it is possible for the homogeneously mixing epidemic model by careful updating of the calculation of the likelihood for the time taken per iteration to be smaller for smaller $k$. In such cases the optimal acceptance rate will be larger than $23.4\%$ and if the time per iteration is proportional to $k$ it will be optimal to update a single component at a time. Secondly, the theoretical results of Section \ref{S:Theo} for independent and identically distributed product densities are shown to give clear guidance for optimising the independence sampler for the homogeneously mixing epidemic model but not for the BDM model. The reason for this difference is not immediately obvious but is likely to depend on the relationship of the observed data to the augmented data. For the homogeneously mixing epidemics the local behaviour of $\mathbf{I}$ is important, for example ensuring $\mathbf{I}$ is consistent with an epidemic outbreak, whereas for the BDM model it is global properties of $(\mathbf{U}, \mathbf{W})$, the total numbers of births, deaths and mutations which are most important. For the random walk Metropolis algorithm optimal scaling results differ depending upon whether the acceptance probability depends on local behaviour (discontinuous product densities, \cite{NRY12}) or global behaviour (continuous product densities, \cite{RGG97}, elliptically
symmetric densities \cite{SR09})  of the proposed moves.

\section*{Acknowledgements}

This research was supported by the  Engineering and Physical Sciences
Research Council under grant EP/J008443/1. We would like to thank an anonymous referee for their careful reading of the paper and suggestions for improving presentation of the findings.


\appendix

\section{Proof of Lemma \ref{Lemma1}} \label{app:A}

Since $\mathbf{Z}_0^n \sim \pi_n$, for all $0 \leq s \leq t$, $\mathbf{Z}_s^n \sim \pi_n$, since $\pi_n$ is the stationary distribution of $\mathbf{Z}_t^n$.  Therefore, we have that
\begin{eqnarray} \label{eq:lem1:1}
\pz ( \mathbf{Z}_s^n \not\in \mathcal{A}_n, \mbox{for some } 0 \leq s \leq t) \leq t n \pz (\mathbf{X}_0^n \not\in \mathcal{A}_n).
\end{eqnarray}

Now,
\begin{eqnarray} \label{eq:lem1a:2}
\pz (\mathbf{X}_0^n \not\in \mathcal{A}_n) &=& \pz \left( \int | H (y, \mathbf{X}_0^n) - H^\ast (y,X_{0,1})| q(y) \,dy > n^{-\frac{1}{8}}\right)
\nonumber \\
&=& \int \pz\left( \int | H (y, \mathbf{x}^n) - H^\ast (y,x_1)| q(y) \,dy > n^{-\frac{1}{8}}\right) \pi_n (\mathbf{x}^n) \, d \mathbf{x}^n. \nonumber \\
\end{eqnarray}
Applying Markov's inequality to the right hand side of \eqref{eq:lem1a:2}, we have that
\begin{eqnarray} \label{eq:lem1a:3}
\pz (\mathbf{X}_0^n \not\in \mathcal{A}_n) & \leq& \int \sqrt{n} \left\{ \int | H (y, \mathbf{x}^n) - H^\ast (y,x_1)| q(y) \,dy \right\}^4 \pi_n (\mathbf{x}^n) \, d \mathbf{x}^n. \nonumber \\
\end{eqnarray}
It then follows by Jensen's inequality that
\begin{eqnarray} \label{eq:lem1a:4}
\pz (\mathbf{X}_0^n \not\in \mathcal{A}_n) & \leq& \int \sqrt{n} \left\{ \int ( H (y, \mathbf{x}^n) - H^\ast (y,x_1))^4 q(y) \,dy \right\} \pi_n (\mathbf{x}^n) \, d \mathbf{x}^n \nonumber \\
& = & \sqrt{n} \int \left\{ \int (H (y, \mathbf{x}^n) - H^\ast (y,x_1))^4 \pi_n (\mathbf{x}^n) \, d \mathbf{x}^n  \right\} q(y) \, dy. \nonumber \\
\end{eqnarray}

We now focus on the inner integral on the right hand side of \eqref{eq:lem1a:4}. Since $\ez_{\mathbf{X}^{n-}} [ H (y, x_1, \mathbf{X}^{n-})] = H^\ast (y, x_1)$,  we have that
\begin{eqnarray} \label{eq:lem1a:5}
&& \int (H (y, \mathbf{x}^n) - H^\ast (y,x_1))^4 \pi_n (\mathbf{x}^n) \, d \mathbf{x}^n \nonumber \\
&=& \int \ez [ (H (y, x_1, \mathbf{X}_0^{n-}) - \ez_{\mathbf{X}_0^{n-}} [ H (y, x_1, \mathbf{X}_0^{n-})])^4] f(x_1) \, d x_1.
\end{eqnarray}

Let $\mathcal{I}_n = \{ \mathbf{i} \in \{2,3, \ldots,n\}^{k-1}; i_1 < i_2 <\ldots <i_{k-1} \}$. Then letting
\begin{eqnarray} \label{eq:lem1a:6}
\hat{H}_{\mathbf{i}} (y, x_1, \mathbf{x}^{n-}) = \ez_{\mathbf{Y}^n} \left[ 1 \wedge \frac{\omega (Y_1)}{\omega (x_1)} \prod_{l=1}^{k-1} \frac{\omega (Y_{i_l})}{\omega (x_{i_l})} \right], \end{eqnarray}
we note that for all $\mathbf{i}, \mathbf{j} \in \mathcal{I}_n$,  $\hat{H}_{\mathbf{i}} (y, x_1, \mathbf{X}_0^{n-}) \eqd \hat{H}_{\mathbf{i}} (y, x_1, \mathbf{X}_0^{n-})$, where $\eqd$ denotes equality in distribution. Hence for all $\mathbf{i} \in \mathcal{I}_n$, $\ez [\hat{H}_{\mathbf{i}} (y, x_1, \mathbf{X}_0^{n-})] = H^\ast (y, x_1)$. Therefore given that
\begin{eqnarray} \label{eq:lem1a:7}
H (y,x_1, \mathbf{X}_0^{n-}) = \binom{n-1}{k-1}^{-1} \sum_{\mathbf{i}} \hat{H}_{\mathbf{i}} (y,x_1, \mathbf{X}_0^{n-}),
\end{eqnarray}
it follows  that
\begin{eqnarray} \label{eq:lem1a:8}
& & \ez [ (H (y, x_1, \mathbf{X}_0^{n-}) - \ez_{\mathbf{X}_0^{n-}} [ H (y, x_1, \mathbf{X}_0^{n-})])^4] \nonumber \\
&=& \binom{n-1}{k-1}^{-4} \sum_{\mathbf{i}_1 \in \mathcal{I}_n} \sum_{\mathbf{i}_2 \in \mathcal{I}_n} \sum_{\mathbf{i}_3 \in \mathcal{I}_n} \sum_{\mathbf{i}_4 \in \mathcal{I}_n} \ez \left[ \prod_{j=1}^4 (\hat{H}_{\mathbf{i}_j} (y,x_1, \mathbf{X}_0^{n-}) - \ez [\hat{H}_{\mathbf{i}_j} (y,x_1, \mathbf{X}_0^{n-}) ] )  \right]. \nonumber \\
\end{eqnarray}
Note that if $\mathbf{i}, \mathbf{j} \in \mathcal{I}_n$ have no elements in common then $\hat{H}_{\mathbf{i}} (y,x_1, \mathbf{X}_0^{n-})$ and $\hat{H}_{\mathbf{j}} (y,x_1, \mathbf{X}_0^{n-})$ are independent. Therefore $\ez [ \prod_{j=1}^4 (\hat{H}_{\mathbf{i}_j} (y,x_1, \mathbf{X}_0^{n-}) - \ez [\hat{H}_{\mathbf{i}_j} (y,x_1, \mathbf{X}_0^{n-}) ] )  ]$ is only non-zero if and only if for $j =1,2,3,4$, $\mathbf{i}_j$ has at least an element in common with one the other indices.
Moreover, $|\ez [ \prod_{j=1}^4 (\hat{H}_{\mathbf{i}_j} (y,x_1, \mathbf{X}_0^{n-}) - \ez [\hat{H}_{\mathbf{i}_j} (y,x_1, \mathbf{X}_0^{n-}) ] )  ]| \leq 1$.

The number of combinations of $\mathbf{i}_1, \mathbf{i}_2 \in \mathcal{I}_n$ such that $\mathbf{i}_1$ and $\mathbf{i}_2$ have at least one element in common is
\begin{eqnarray} \label{eq:lem1a:9} \binom{n-1}{k-1} \left\{  \binom{n-1}{k-1} -  \binom{n-k}{k-1} \right\}, \end{eqnarray} which is bounded above by $n^{2k-3}/\{ (k-2)!\}^2$ for all sufficiently large $n$. Similarly, the number of combinations of $\mathbf{i}_1, \mathbf{i}_2, \mathcal{i}_3, \mathcal{i}_4 \in \mathcal{I}_n$ such that $\mathbf{i}_2$, $\mathbf{i}_3$ and $\mathbf{i}_4$ all have at least one element in common with $\mathbf{i}_1$ is
\begin{eqnarray} \label{eq:lem1a:10} \binom{n-1}{k-1} \left\{  \binom{n-1}{k-1} -  \binom{n-k}{k-1} \right\}^3, \end{eqnarray} which is bounded above by $(k-1)^2 n^{4k-7}/\{ (k-2)!\}^4$ for all sufficiently large $n$. Now $\ez [ \prod_{j=1}^4 (\hat{H}_{\mathbf{i}_j} (y,x_1, \mathbf{X}_0^{n-}) - \ez [\hat{H}_{\mathbf{i}_j} (y,x_1, \mathbf{X}_0^{n-}) ] )  ]$ is only non-zero if either $\mathbf{i}_1, \mathbf{i}_2, \mathcal{i}_3, \mathcal{i}_4 \in \mathcal{I}_n$ can be grouped into two pairs such that both pairs have at least one element in common or if three of the components all have at least one element in common with the fourth. (Note that there is overlap between these two classifications.) Thus using \eqref{eq:lem1a:9} and \eqref{eq:lem1a:10}, it is straightforward to combine with \eqref{eq:lem1a:8} to show that
\begin{eqnarray} \label{eq:lem1a:11} && \ez [ (H (y, x_1, \mathbf{X}_0^{n-}) - \ez_{\mathbf{X}_0^{n-}} [ H (y, x_1, \mathbf{X}_0^{n-})])^4] \nonumber \\ & \leq & \binom{n-1}{k-1}^{-4} \left\{ 3 \left(\frac{n^{2k-3}}{\{(k-2)!\}^2} \right)^2 + 4 \frac{(k-1)^2 n^{4k-7}}{\{ (k-2)!\}^4} \right\} \nonumber \\ & \leq & \frac{(k-1)^4}{(n-k)^{4k-4}} \left\{ 3 n^{4k-6} + 4 (k-1)^2 n^{4k-7} \right\}. \end{eqnarray}
Since the bound obtained in \eqref{eq:lem1a:11}, holds for all $y, x_1 \in \mathbb{R}$, it follows from \eqref{eq:lem1a:4} and  \eqref{eq:lem1a:5} that
\begin{eqnarray} \label{eq:lem1a:12}
n \pz (\mathbf{X}_0^n \not\in \mathcal{A}_n) & \leq& n \sqrt{n} \frac{(k-1)^4}{(n-k)^{4k-4}} \left\{ 3 n^{4k-6} + 4 (k-1)^2 n^{4k-7} \right\} \nonumber \\
& \rightarrow & 0 \hspace{0.5cm} \mbox{as } \nr. \end{eqnarray} The lemma immediately follows by combining \eqref{eq:lem1a:12} and \eqref{eq:lem1:1}.

\end{document}